\documentclass[10pt,twocolumn,letterpaper]{article}

\usepackage{wacv}              
\usepackage{graphicx}
\usepackage{amsmath}
\usepackage{amssymb}
\usepackage{booktabs}
\usepackage{makecell}  
\usepackage{multirow}
\usepackage{textcomp} 

\usepackage[pagebackref,breaklinks,colorlinks]{hyperref}

\usepackage[capitalize]{cleveref}
\crefname{section}{Sec.}{Secs.}
\Crefname{section}{Section}{Sections}
\Crefname{table}{Table}{Tables}
\crefname{table}{Tab.}{Tabs.}

\def\wacvPaperID{907} 
\def\confName{WACV}
\def\confYear{2025}

\begin{document}

\title{Context-Aware Optimal Transport Learning for Retinal Fundus Image Enhancement}

\author{Vamsi Krishna Vasa$^{*}$\\
Arizona State University\\
{\tt\small vvasa1@asu.edu}
\and
Peijie Qiu$^{*}$\\
Washington University in St.Louis\\
{\tt\small peijie.qiu@wustl.edu}
\and
Wenhui Zhu \\
Arizona State University\\
{\tt\small wzhu59@asu.edu}
\and
Yujian Xiong \\
Arizona State University\\
{\tt\small yxiong42@asu.edu }
\and
Oana Dumitrascu \\
Mayo Clinic\\
{\tt\small dumitrascu.oana@mayo.edu }
\and
Yalin Wang \\
Arizona State University\\
{\tt\small ylwang@asu.edu }
}
\maketitle
\begin{abstract}
   Retinal fundus photography offers a non-invasive way to diagnose and monitor a variety of retinal diseases, but is prone to inherent quality glitches arising from systemic imperfections or operator/patient-related factors. However, high-quality retinal images are crucial for carrying out accurate diagnoses and automated analyses. The fundus image enhancement is typically formulated as a distribution alignment problem, by finding a one-to-one mapping between a low-quality image and its high-quality counterpart. This paper proposes a context-informed optimal transport (OT) learning framework for tackling unpaired fundus image enhancement. In contrast to standard generative image enhancement methods, which struggle with handling contextual information  (e.g., over-tampered local structures and unwanted artifacts), the proposed context-aware OT learning paradigm better preserves local structures and minimizes unwanted artifacts. Leveraging deep contextual features, we derive the proposed context-aware OT using the earth mover's distance and show that the proposed context-OT has a solid theoretical guarantee. Experimental results on a large-scale dataset demonstrate the superiority of the proposed method over several state-of-the-art supervised and unsupervised methods in terms of signal-to-noise ratio, structural similarity index, as well as two downstream tasks. The code is available at \url{https://github.com/Retinal-Research/Contextual-OT}.
   
\end{abstract}

\def\thefootnote{*}\footnotetext{These authors contributed equally to this paper.}

\section{Introduction}
\label{sec:intro}

Retinal color fundus photography (CFP) is vital for diagnosing ocular diseases, with non-mydriatic CFP being increasingly used for point-of-care diagnosis \cite{OANA3}. CFP also plays an indispensable role in screening neurodegenerative disorders, such as Alzheimer's disease and systemic conditions (e.g., diabetes mellitus~\cite{Oana,Wagner:Systmic2020}). However, the quality of non-mydriatic retinal CFPs affected by various factors making it harder for accurate diagnosis in some cases. 


Early Optical models \cite{41493, cheng2018structure} were capable of handling the quality degradation due to opaque internal cataractous media and yielding high-quality counterparts. However, the CFP is highly influenced by the manual errors arising from the imaging equipment and environmental conditions (e.g., dim surroundings without sufficient lights). The quality of fundus photography exhibits significant variability attributed to multiple factors, including varying operator expertise, fluctuations in illumination during image capture, lens contamination, and abrupt adjustments to focus settings resulting in blurred images. These noises compromise image quality and obscure crucial details such as blood vessels and lesions. Developing a single technique to robustly improve low-quality CFPs suffering from the aforementioned factors would aid in disease (e.g., diabetic retinopathy) diagnosis as well as develop automated tools for screening and population studies of neurological disorders. 


Recently, deep learning based methods have achieved state-of-the-art performance in enhancing the quality of fundus images. Early explorations in retinal fundus image enhancement revolve around supervised learning~\cite {shen2020modeling,10.1007/978-3-031-16434-7_49,li2023generic}, which requires the noisy-clean pairs. However, the collection of paired noisy-clean retinal training samples proved arduous and costly in practice. To mitigate this challenge, unsupervised methods such as Generative Adversarial Networks (GANs) \cite{goodfellow2014generative} have drawn significant attention in recent years by modeling fundus image enhancement as an image translation task~\cite{9763342,zhu2023optimal,zhu2023otre}. One notable work is the OT-based generative models for fundus image enhancement~\cite{zhu2023optimal,zhu2023otre}, which leverages the fact that low-quality images and their high-quality counterparts share similar underlying structures. This helps reduce the search space for unpaired image-to-image tasks without a computationally expensive cycle consistency~\cite{cyclegan}.



Another major challenge towards robust enhancement is understanding the contextual differences between the high-quality and poor-quality domains. Most of the latest enhancement techniques rely on a quadratic cost or an SSIM cost for the preservation of delicate details such as lesions and blood vessels.  While this helps preserve structural information, it also distills contextually unwanted artifacts (e.g., light spots) to the enhanced images. It arises due to the fundamental limitation of SSIM in overlooking contextual information in image enhancement. Drawing on the understanding that contextual information is often embedded in the deep layers of pre-trained neural networks~\cite{gatys2016image,mechrez2018contextual}, our approach shifts the computation of the OT cost from image space to embedding space. Our innovative context-aware OT framework leverages deep feature spaces for more accurate fundus image enhancement, supported by the theoretical foundations of earth mover's distance and OT theory, thus providing robust theoretical underpinnings. 

Our main contributions are threefold: \textbf{(i)} We introduce a novel OT retinal image enhancement learning paradigm based on the deep layer feature space, aiming to minimize undue excessive tampering to lesions and structures while effectively removing noise. 
\textbf{(ii)} Our method offers a strong theoretical foundation for general image enhancement tasks by ensuring that the transport cost reflects the intrinsic geometrical and contextual properties of the data in the deep feature space. 
\textbf{(iii)} Our comprehensive evaluation across three large publicly available retinal imaging datasets demonstrated the superiority of the proposed method over strong unsupervised and supervised competing methods.

\section{Related Work}
Recent advancement of deep learning has achieved state-of-the-art performance on the fundus image enhancement task. Prior deep learning based methods for fundus image enhancement can be roughly divided into three categories: (i) supervised methods and , (ii) self-supervised methods, and (iii) unsupervised methods. The supervised methods~\cite{shen2020modeling,10.1007/978-3-031-16434-7_49} have a hard requirement on paired noisy-clean images, while the unsupervised methods~\cite{cyclegan,arcnet,9763342,zhu2023optimal,zhu2023otre} are typically trained on unpaired dataset. The self-supervised methods rely on the information of the training dataset itself to formulate a supervised learning scheme. 

Focusing on the supervised methods, Shen et al. \cite{shen2020modeling} introduces a clinically-focused fundus enhancement network (cofe-Net) that learns a direct mapping from degraded noisy images to high-quality clean images in a supervised fashion. Specifically, this approach leverages early-stage low-quality region activation and continuous retinal structure injection via a cascaded encoder-decoder network at multiple scales with shared weights. Recently, Liu et al. \cite{10.1007/978-3-031-16434-7_49} proposes PCE-Net to decompose low-quality images into Laplacian pyramid features for multi-resolution based enhancement. The added feature pyramid constraint for the sequence guides the PCE-Net to be degradation-invariant. However, these methods have limitations in real practice due to their reliance on paired noisy-clean images.

To relax the requirement of paired training samples, Li et al. \cite{li2023generic} introduces a fundus image enhancement network boosted by frequency self-supervised representation learning with structure-aware enhancement. This approach combines frequency self-supervision and synthesized data to train GFE-Net. Following this vein, SCR-Net \cite{scrnet} introduces an enhancement network, which involves synthesizing multiple cataract-affected images from a clear fundus image followed by aligning and restoring high-frequency components. SCR-Net comprises an encoder for capturing high-frequency components and a decoder for enforcing high-frequency  alignment to facilitate structure alignment and fundus image enhancement through feature sharing.

\begin{figure*}[!t]
  \centering
  \includegraphics[width=\textwidth]{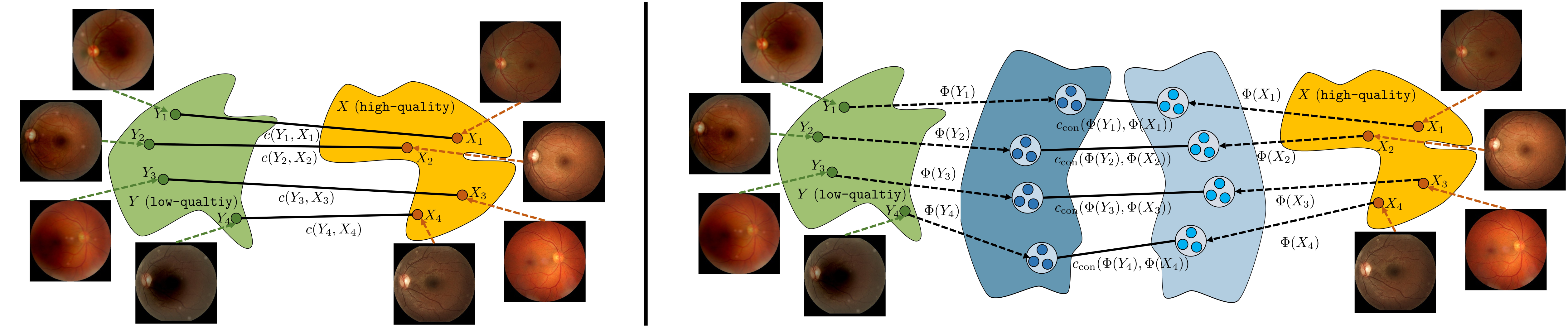}  
  \caption{Comparison between the traditional OT learning scheme (\textbf{Left}) and the proposed context-aware OT scheme (\textbf{Right}) for fundus image enhancement. Different from the traditional OT learning scheme that performs OT on image space, our contextual OT performs OT on the contextual feature space, which can help preserve contextual information between low-quality and high-quality images.}
  \label{Context-aware-OT}
\end{figure*}

Due to the difficulty of collecting noisy-clean retinal image pairs, unsupervised methods such as GANs \cite{goodfellow2014generative} have drawn significant attention in recent years by modeling fundus image enhancement as an image-to-image translation task. In particular,  CycleGAN~\cite{cyclegan} serves as the most common image translation-based method for fundus image enhancement by training on unpaired low-quality and high-quality fundus images. Arguably, it suffers from a large search space with a huge computational overhead as well as the failure to preserve vessel and lesion structures. To address this limitation, ArcNet \cite{arcnet} utilizes multiple quadratic loss functions to enforce the alignment of high-frequency components between low-high quality images. 
I-SECRET \cite{i-secret} introduces a dual-stage approach, including a supervised learning stage trained on degraded-clean image pairs and an unsupervised learning stage that focuses on generalizing enhancements using GAN. NAGAN \cite{8852672} introduces an approach that focuses on learning speckle noise patterns for OCT image-to-image translation. This is achieved by using a generator that takes images from the source domain as input and produces output images with noise resembling that of the target domain. Two discriminators are then utilized: one ensures that the generated images replicate the noise patterns of the target domain, while the other ensures that the structures from the source domain remain intact. However, this method is specifically designed for OCT image translation, where the primary distinction between the source and target domains lies in the speckle noise patterns.


In addition, GANs \cite{9763342,zhu2023optimal,zhu2023otre}  that leverage optimal transport (OT) theory to reduce search space have also been explored. These methods 
are contigent on the fact that low-quality images and their high-quality counterparts should share the same underlying structures. Wang et al.~\cite{9763342} propose an OT guided GAN (OTTGAN) for unsupervised image denoising with a single generator and discriminator. Although it achieved significant results in natural image denoising, its adoption of a quadratic OT cost led to the destruction or over-tampering of the vessel and lesion structures. To address these challenges, Zhu et al. introduce OTGAN~\cite{zhu2023otre} and OTEGAN~\cite{zhu2023optimal}, which leverage a structural similarity index (SSIM) cost to preserve structural information (e.g., lesions, vessel structures, and optical discs) between enhanced and low-quality images. It is worth noting that OTEGAN~\cite{zhu2023otre} is an extension of OTGAN~\cite{zhu2023optimal}, with an additional post-preprocessing step termed regularization by enhancing.
While this helps preserve structural information, SSIM also distills contextually unwanted artifacts (e.g., light spots) to the enhanced images. It arises due to the fundamental limitation of SSIM in overlooking contextual information in image enhancement, as it operates on image space.
\section{Method}

\noindent \textbf{Monge's formulation.} The fundus image enhancement task is formulated as an unpaired image-to-image translation task. Specifically, given the source domain $Y$ (low-quality domain) to the target domain $X$ (high-quality domain), our goal is to find a direct transformation $f: Y \rightarrow X$. The idea of applying OT to this problem is natural, as we assume that there is a one-to-one mapping between low-quality images and high-quality images. The image enhancement problem can be defined by Monge's OT formulation as in~\cite{zhu2023otre}. Specifically, for two probability measures $\nu \sim \mathcal{P}(X)$ and $\mu \sim \mathcal{P}(Y)$, this is given as 

\begin{equation}\label{eq:1}
\min_{f} \left\{\inf \int_{Y} c(y, f(y)) d \nu(y) \right \} \text{subject to} \ \mu = f_{\#} \nu,
\end{equation}

\noindent where $c(\cdot, \cdot)$ is a cost function. Commonly used cost functions are linear cost and quadratic cost. More generally, $c(\cdot, \cdot)$ can be defined as $c(|x-y|)$ for some convex function $c$ that measures the discrepancy between $x$ and $y$. It is worth noting that each image is treated as a point in its support, i.e., $x \sim X$ and $y \sim Y$. The resulting OT learning scheme for image enhancement is shown in Fig.~\ref{Context-aware-OT} (\textbf{Left}).

\noindent \textbf{Lagrangian relaxation.}\label{lang-relax} With a Lagrangian multiplier, Eq. (\ref{eq:1}) can be reformulated as an unconstrained optimization problem~\cite{9763342,zhu2023otre}:
\begin{equation}\label{eq:2}
\begin{split}
     \min_{f} & \ \underbrace{\mathbb{E}_{Y \sim P_{Y}} c(Y,  f(Y))}_{\text{transport cost}} + \underbrace{\lambda d(p_{\hat{X}}, p_{X})}_{\text{divergence}}, 
\end{split}
\end{equation}
where $d(\cdot, \cdot)$ measures the divergence between probability distribution $p_{\hat{X}}$ and $p_X$. Here, we use $\hat{X} = f(Y)$ to define the enhanced domain. The first term in Eq.~(\ref{eq:2}) minimizes the transport cost from the low-quality domain to the high-quality domain, facilitating maximal preservation of information from low-quality images in the enhanced images. The second term aligns the enhanced domain with the high-quality domain distribution. Notably, this formulation does not require cycle consistency \cite{cyclegan}, ensuring computational efficiency. It also leverages prior knowledge that enhanced/high-quality images should share underlying structures (e.g., optic disc, lesions, vessels) with low-quality images but are degraded by factors like illumination pollution, retinal artifacts, and blurring \cite{shen2020modeling}. This can reduce the search space in cycle consistency and mitigate the introduction of unrealistic components from CycleGAN.

\begin{figure*}[!t]
  \centering
  \includegraphics[width=0.9\textwidth]{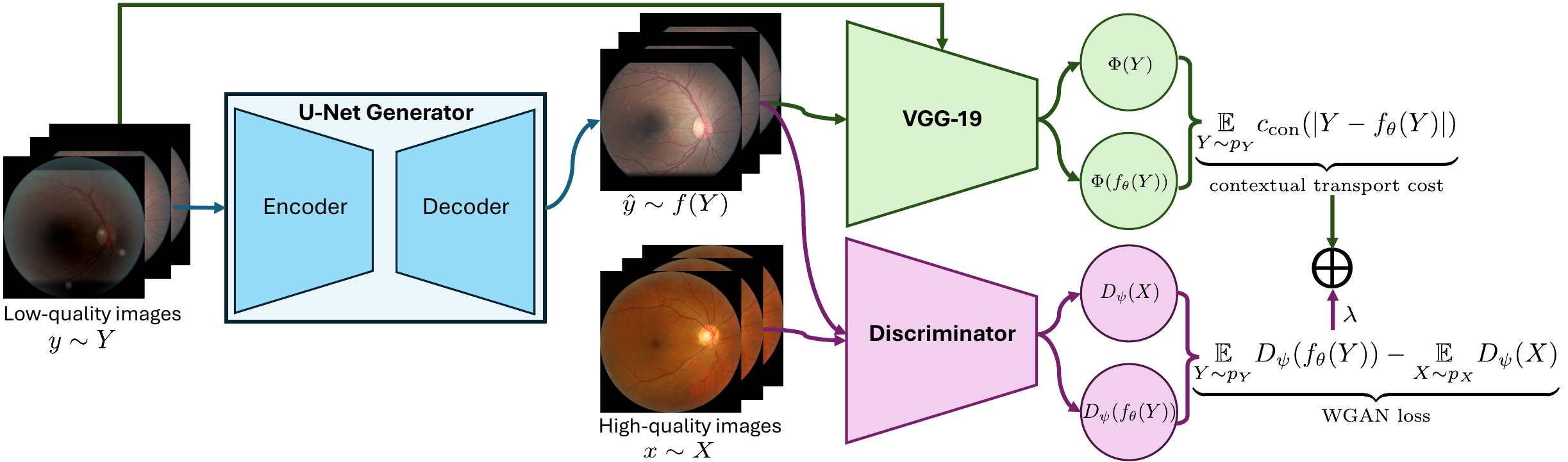}  
  \caption{The adversarial training scheme of the proposed contextual OT. The generator (\textit{$f_{\theta}$}) is a U-Net with residual connection and channel attention as outlined in~\cite{zhu2023otre,9763342}. The discriminator ($\Phi$) is also the one used in~\cite{zhu2023otre,9763342}. We use a VGG-19 for encoding the contextual information onto feature space and compute the contextual transport cost based on these feature embeddings.}
  \label{network}
\end{figure*}


\subsection{Context-Aware OT for Image Enhancement}\label{sec:cot}
We argue that the problem defined in Eq. (\ref{eq:2}) is suboptimal for the image enhancement task. This is largely due to the fact that the common choice of the transport cost function $c$ (e.g., linear cost $c(Y, f(Y)) = ||Y - f(Y)||$, quadratic cost $c(Y, f(Y)) = ||Y - f(Y)||^2$ in~\cite{9763342} and SSIM cost in~\cite{zhu2023otre}) can not effectively handle the image contextual information. 
Specifically, the linear/quadratic cost only encourages the enhanced images to have the same arithmetic median/mean as low-quality images. 
However, the linear/quadratic cost treats each pixel independently but cannot effectively preserve structural information. The SSIM cost is only locally quasi-convex, which causes the solution to Eq. (\ref{eq:2}) inherently suboptimal. Although the SSIM cost helps preserve local information, it also retains contextually unwanted retinal artifacts in the enhanced images, as retinal artifacts also exhibit meaningful local structures. 

In a short word, the aforementioned cost functions cannot effectively handle the image context, as they operate in the image space.
Instead, the contextual information is typically abstracted in the deeper layers of a neural network~\cite{gatys2016image,mechrez2018contextual}. Inspired so, we explore incorporating contextual information into the OT problems defined in Eq. (\ref{eq:1}) and (\ref{eq:2}). We derive the context-aware OT using the earth mover's (EM) distance, as we will consider discrete sets instead of a continuous one in the previous derivation. Similar to~\cite{gatys2016image,mechrez2018contextual}, we represent each image $X$ and $Y$ as a collection of feature vectors in the deep layer of a neural network $\Phi$ (e.g., VGG in~\cite{mechrez2018contextual}): $Y=\{\Phi(y_i)\}$ and $f(Y)=\{\Phi(f(y_j))\}$ with $|X| = |Y|=N$. We also normalize the feature vectors to have a unit length $||\Phi(y_i)||^2=1$, $||\Phi(f(y_j))||^2=1$ to make it scale invariant. 
The EM distance~\cite{rubner1998metric} is given as
\begin{equation}
\begin{split}
    & \quad d_{\text{EM}}(Y, f(Y)) = \min_{F \geq 0} \sum_{ij} F_{ij} C_{ij} \\
    & \text{subject to} \quad \sum_j F_{ij} = \sum_i F_{ij} = \frac{1}{N}.
\end{split}
\end{equation}

\noindent where $F$ is the flow matrix, and $C$ is the cost matrix. Due to its computational intractability, we consider relaxed EM (REM) distance~\cite{kusner2015word}, which can be easily  optimized via gradient descent. Formally, this is given as
\begin{equation}\label{eq:4}
\begin{split}
     d_{\text{REM}}(Y, f(Y)) &= \max\{\frac{1}{N}\sum_{i} \min_j C_{ij}, \frac{1}{N}\sum_{j} \min_i C_{ij} \}.
\end{split}
\end{equation}
According to the OT property, the transport map is symmetric from $X$ to $Y$ and $Y$ to $X$ (i.e., with the same cost)~\cite{symmetric}, which is valid for our image enhancing task, we can remove the maximum operation in Eq. (\ref{eq:4}).
To this end, we formally derive our context-aware OT cost as
\begin{equation}
    c_{\text{con}}(|Y - f(Y)|) = \frac{1}{N}\sum_{j} \min_i C_{ij}.
\end{equation}
The yielded context-aware OT learning scheme is shown in Fig.\ref{Context-aware-OT} (\textbf{Right}). Now, we consider the convex function w.r.t. distance to obtain the cost matrix $C$ (e.g., euclidean distance). We also consider the distance defined in~\cite{mechrez2018contextual}:
\begin{equation}
    C_{ij} = \exp\left(\frac{{||\Phi(y_i) - \Phi(f(y_j)) ||^2}}{h} \right),
\end{equation}
where $h > 0$ is a smoothing band-width parameter; when $h=0.5$, the above cost converges to a cosine similarity-based cost.



\noindent \textbf{Context-aware OT learning as a GAN.}\label{C-OTGAN}
For simplicity, the optimization problem defined in Eq.~(\ref{eq:2}) can be realized as a GAN with a Wasserstein distance (see Fig.~\ref{network}). Following the convention of the WGAN, we can derive the adversarial training scheme:
\begin{align}
\label{loss_eq}
    &\min_{\substack{ ||D_{\psi}||_L \leq 1}} \left\{ \underbrace{\underset{Y \sim p_{Y}}{\mathbb{E}} c_{\text{con}}(|Y-f_{\theta}(Y)|)}_{\text{contextual transport cost}} \right\} \nonumber \\
    &\qquad + \lambda \left\{ \underbrace{\underset{Y \sim p_{Y}}{\mathbb{E}} D_{\psi}(f_{\theta}(Y)) - \underset{X \sim p_{X}}{\mathbb{E}} D_{\psi} (X)}_{\text{WGAN loss}} \right\}.
\end{align}
where $f_{\theta}$, $D_{\psi}$ are generator and discriminator parameterized by $\theta$ and $\psi$, respectively.
It is worth noting that the GAN solution is subjected to the constraint that the function $f_{\psi}$ and $D_{\psi}$ are both $1$-Lipschitz continuous. In our implementation, we use spectral normalization and gradient penalty to impose $1$-Lipschitz constraint to $f_{\theta}$ and $D_{\psi}$, respectively. While other alternative solutions~\cite{gazdieva2022optimal,korotin2022neural} do not require this constraint. We will not discuss them here as they are beyond the scope of this paper. For a fair comparison, we leverage the same generator and discriminator network architectures outlined in~\cite{zhu2023otre}.



\begin{figure*}[!t]
  \centering
  \includegraphics[width=0.8\textwidth]{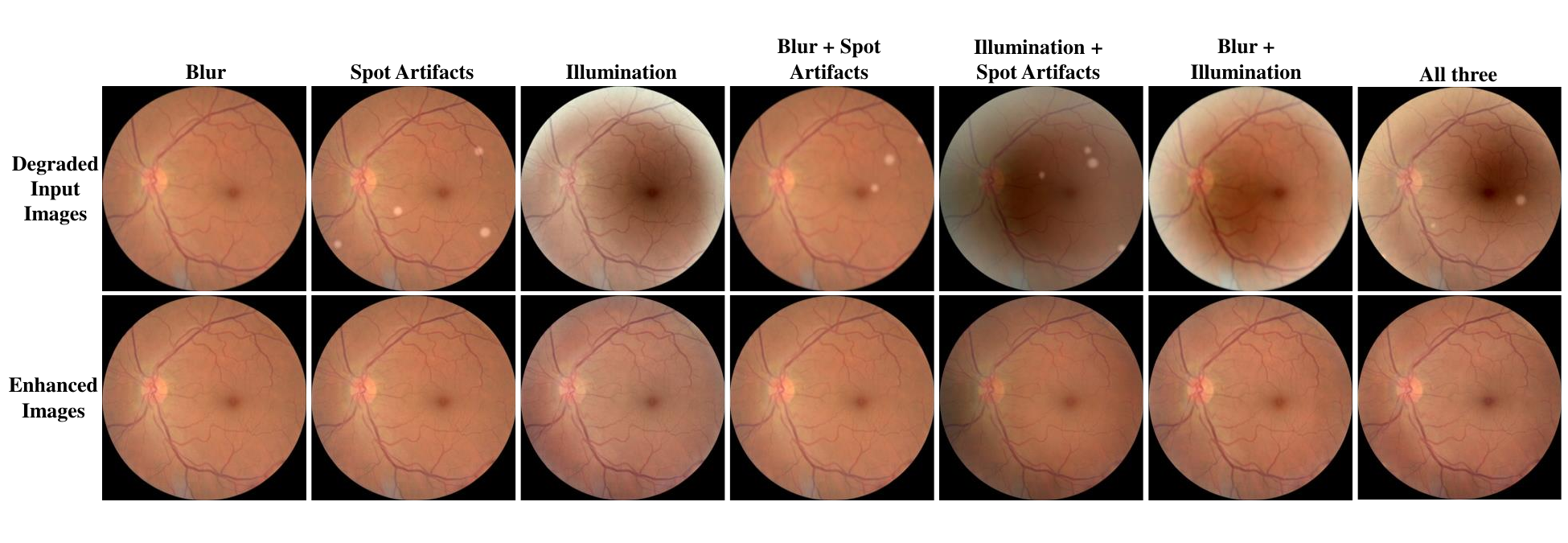}  
  \vspace{-0.4cm}
  \caption{Qualitative results of the proposed method on degraded images over the combinations of different noise (i.e., spot artifacts, illumination, and blurring). Our method achieves good enhancement performance even on severely degraded images (\textbf{cols.} 4, 5, and 6).}
  \label{degrade-perf}
\end{figure*}
\section{Experiments and results}
\label{sec:Experiment}
\subsection{Experimental design}
 We validated the effectiveness of the proposed method on three publicly available datasets: the EyeQ \cite{eyeq}, DRIVE \cite{drive}, and IDRID \cite{idrid}. Following~\cite{zhu2023otre}, we trained the proposed method using the EyeQ dataset and evaluated it on the downstream tasks, i.e vessel segmentation and diabetic lesion segmentation, using DRIVE and IDRID datasets. 

\begin{table*}[!t]
\centering
\caption{Performance comparison with the SOTA methods. The best performance within each column is highlighted in bold. ($^*$: $p<0.01$; with the paired $t$-test to the baseline methods.)}
\tiny
\resizebox{0.9\textwidth}{!}{%
\begin{tabular}{@{}cccccccc@{}}
\toprule
\multirow{2}{*}{} & \multirow{2}{*}{\textbf{Method}} & \multicolumn{2}{c}{\textbf{EyeQ}} & \multicolumn{2}{c}{\textbf{DRIVE}} & \multicolumn{2}{c}{\textbf{IDRID}} \\ \cmidrule(l){3-8} 
                                          &                                  & \textbf{PSNR}   & \textbf{SSIM}   & \textbf{PSNR}    & \textbf{SSIM}   & \textbf{PSNR}    & \textbf{SSIM}   \\ \midrule
\multirow{2}{*}{\textit{Supervised Methods}}       & cofe-Net~\cite{shen2020modeling}                         & 17.25           & 0.880           & 19.11            & 0.66            & 19.07            & 0.65            \\
                                          & PCE-Net~\cite{10.1007/978-3-031-16434-7_49}                          & 18.54           & 0.874           & 28.94            & 0.646           & 20.26            & 0.775           \\ \midrule
\multirow{8}{*}{\textit{Unsupervised Methods}}     & GFE-Net~\cite{li2023generic}                          & 18.68           & 0.80            & 15.14            & \textbf{0.715}  & 19.21            & 0.631           \\
                                          & CycleGAN~\cite{cyclegan}                         & 22.93           & 0.878           & 21.59            & 0.653           & 21.01            & \textbf{0.764}  \\
                                          & SCR-Net~\cite{scrnet}                          & 18.86           & 0.796           & 18.50            & 0.668           & 19.51            & 0.616           \\
                                          & I-SECRET~\cite{i-secret}                         & 14.84           & 0.884           & 18.75            & 0.669           & 18.40            & 0.756           \\
                                            \cmidrule(l){2-8}
                                          & OTTGAN~\cite{9763342}                           & 23.25           & 0.895           & 20.73            & 0.637           & 20.94            & 0.755           \\
                                          & OTEGAN~\cite{zhu2023otre}                           & 23.51           & 0.898           & 17.96            & 0.601           & 18.20            & 0.687           \\ \cmidrule(l){2-8}
                                          & Ours                       & \textbf{24.79\textsuperscript{*}} & \textbf{0.914\textsuperscript{*}} & \textbf{29.47\textsuperscript{*}}  & 0.673\textsuperscript{*}          & \textbf{21.65\textsuperscript{*}}  & 0.757\textsuperscript{*}          \\ \bottomrule
\end{tabular}%
}
\label{deg-exp}
\end{table*}

\noindent \textbf{Degradation Experiment.} The degradation experiment was conducted to observe the effectiveness of the proposed method on the synthetically degraded retinal fundus images. The training set consists of a subset of 3560 high-quality images from the EyeQ dataset selected based on the Grading label provided. We evaluated the trained weights on complete DRIVE and IDRID images, along with the subset of 1819 high-quality images from the EyeQ dataset. We would like to point out that PSNR and SSIM were estimated between the enhanced images of low-quality images, which were generated by degrading the high-quality image using the model outlined in~\cite{shen2020modeling}, and the corresponding high-quality images. We showcased the prowess of the proposed method over the Degradation cases in the combination of Light Transmission Disturbance, Image Blurring, and Retinal Artifact. The Downstream segmentation tasks are conducted to showcase the effective preservation of the intricate details from the low-quality fundus images post-enhancement. 

\noindent \textbf{Downstream Vessel Segmentation.} The vessel segmentation task is conducted using the DRIVE dataset, where annotated masks are available. We use the official training/testing split, which results in 20 subjects in training and testing set . The Vessel Segmentation task is evaluated based on the Area under ROC, Precision-Recall curve, Sensitivity, and Specificity.

\noindent\textbf{Downstream Diabetic Lesion Segmentation.} We choose the segmentation masks provided with the IDRID dataset. Since the training and testing of downstream segmentation tasks were based entirely on enhanced images, without adding any preprocessing and additional tricks, we only considered large blocks of lesions that were easy to train, such as EX and HE. The Training set is made of 54 subjects, and the Testing set is made of 27 subjects. The performance is quantified using Area under ROC, Precision-recall, and Jaccard Index. We leveraged the vanilla UNet model to train from scratch for both segmentation tasks. 

All images underwent center-cropping and resizing to a dimension of 256 × 256. We compared the proposed method with previous works in different training schemes. Same are as followed, \textit{Supervised methods}: cofe-Net \cite{shen2020modeling}, PCE-Net \cite{10.1007/978-3-031-16434-7_49}, \textit{Unsupervised or GAN based methods}: GFE-Net \cite{li2023generic}, CycleGAN \cite{cyclegan}, SCR-Net \cite{scrnet}, I-SECRET \cite{i-secret}, \textit{OT based techniques}: OTTGAN \cite{9763342} and OTEGAN \cite{zhu2023otre}

\noindent\textbf{Implementation details.} \label{imp-details}
For the Degradation Experiment, We trained the model for 100 epochs using an RMSprop optimizer. The batch size was set to 2. The initial learning rate was set to be \(1 \times 10^{-4}\) for the discriminator and \(5 \times 10^{-5}\) for the generator. The learning rate decayed by a factor of 10 every 50 epochs. To prevent overfitting during training, we employed data augmentation for training, such as random horizontal/vertical flips, random crops, and random rotations. 
The UNet \cite{ronneberger2015unet} as a backbone is leveraged for the Lesion Segmentation task. The model was trained on the summation of BCELoss and Dice loss. We utilized Adam optimizer with the initial learning rate of \(2 \times 10^{-4}\) along with the weight decay set as \(5 \times 10^{-4}\). We maintained the batch size of 4 and trained for 300 epochs. We incorporated the following data augmentation traits: Horizontal and  Vertical Flip, Random Grid Shuffle, and coarse dropout with a probability of 0.5. 
Similar to Lesion segmentation, we used UNet for the Vessel segmentation task on the DRIVE \cite{drive} dataset. We implemented the Adam optimizer with the criterion as Cross Entropy loss, with the initial learning rate as \(5 \times 10^{-5}\) and batch size set to 64. The best model was saved from 50 epochs. To overcome the limited dataset issues, we used the following data augmentation techniques: Random Crop, Random flip (Left-Right and Up-Down) with a probability of 0.5, and Random rotation. All the mentioned experiments were performed on an Nvidia RTX3090 GPU.

\begin{figure}[h]
  \centering
  \includegraphics[width=\columnwidth]{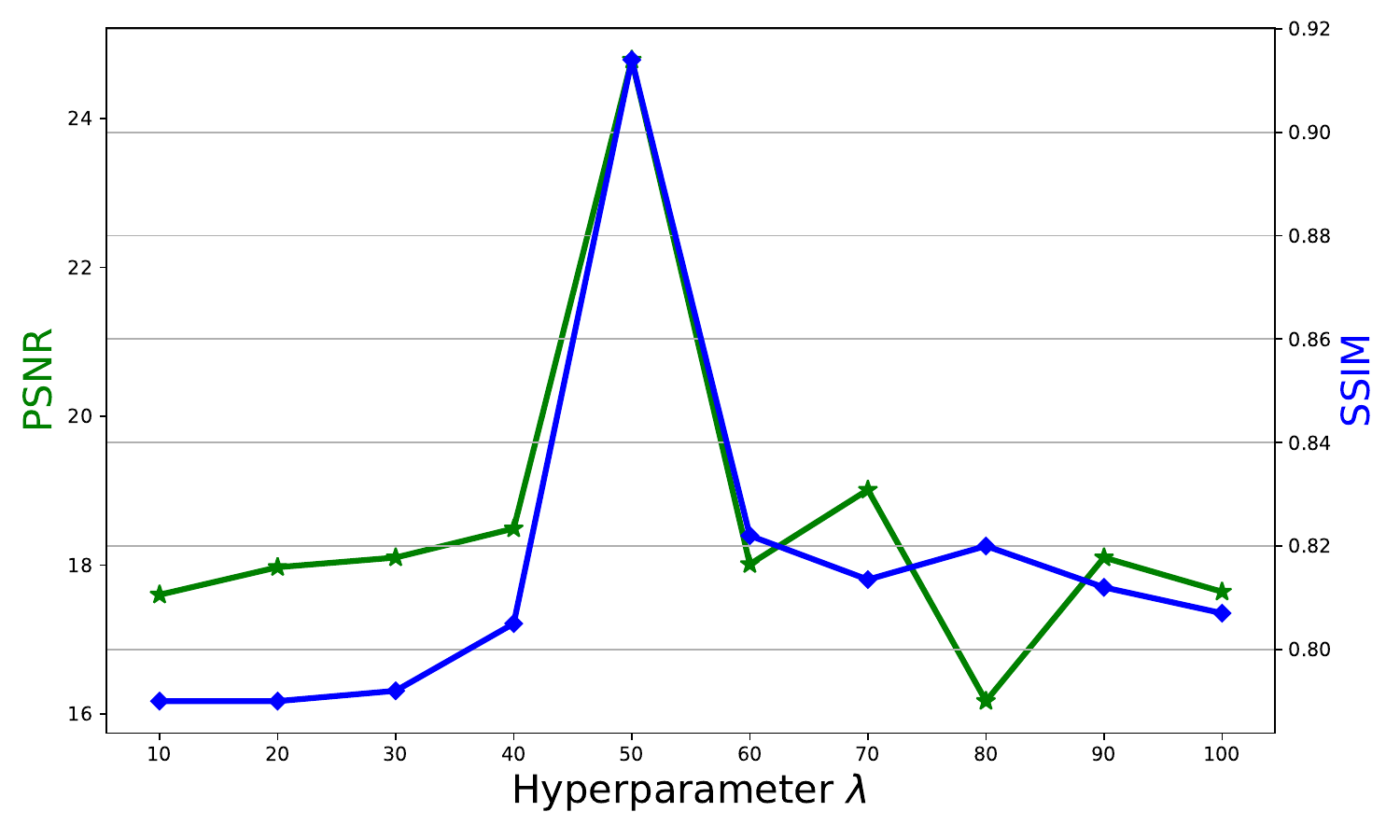}  
  \caption{Ablation study on different $\lambda$.}
  \label{ablation study}
\end{figure}

\begin{figure*}[!ht]
  \centering
  \includegraphics[width=0.97\textwidth]{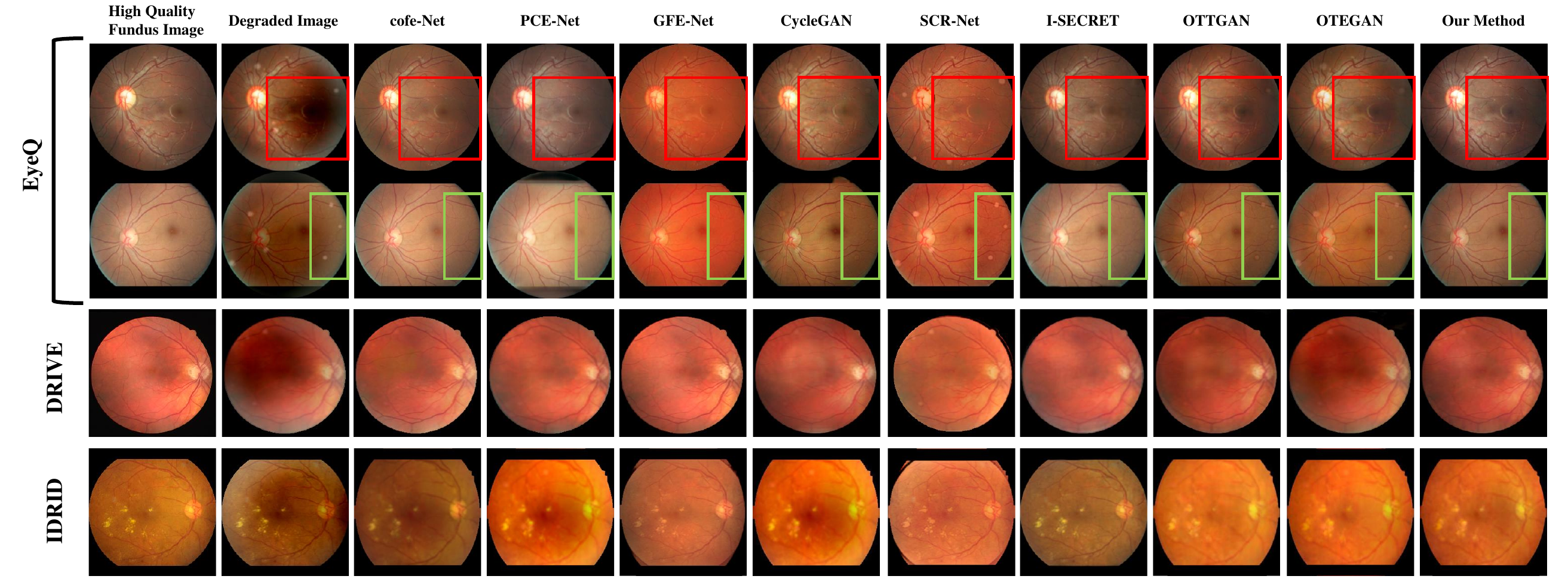}  
  \caption{Visual comparison of our method with baseline methods. \textcolor{red}{Red} box highlights Low Illumination introduced, and \textcolor{green}{Green} box highlights the light spot artifact noise.}
  \label{degrade-exp}
\end{figure*}

\subsection{Ablation studies}

We dedicate this subsection to discussing the ablation study conducted. The Eq.~(\ref{loss_eq}) explains the cost function our approach utilizes.
To understand the significance of the contextual transport cost, We studied the PSNR and SSIM trends over the varying importance of contextual loss in training the Generator. We introduced the multiplication factor $\lambda$ to regulate the importance of Contextual loss. The $\lambda$ covers a wide range from 10 to 100 with a stride of 10. Apart from $\lambda$, the remaining Hyperparameters and training loop are set to the values discussed in \ref{imp-details} for the degradation experiment. We illustrated the performance with respect to the different values of $\lambda$ in Fig. \ref{ablation study}. We observe a slow upward movement of PSNR and SSIM with the increasing $\lambda$ value. The performance is peaked when $\lambda = 50$ (see Fig. \ref{ablation study}), after which, increasing the value of $\lambda$ leads to inferior performance. We hypothesize this might be attributed to the instability of GAN training and the inherent trade-offs between distribution alignment and structure preserving.  


\begin{table*}[!t]
    \centering
    \caption{Performance evaluation of blood vessel and diabetic lesions (EX and HE) segmentation on the DRIVE~\cite{drive} and the IDRID dataset~\cite{idrid}.}
    \tiny
    \resizebox{0.9\textwidth}{!}{%
    \begin{tabular}{lcccc|ccc|ccc}
        \toprule
         \multirow{2}[3]{*}{} & \multicolumn{4}{c}{Vessel Segmentation} & \multicolumn{3}{c}{EX} & \multicolumn{3}{c}{HE} \\ 
         \cmidrule(lr){2-5}  \cmidrule(lr){6-8}  \cmidrule(lr){9-11}
          
         Method & ROC & PR & SE & SP & ROC & PR & Jaccard & ROC & PR & Jaccard\\ \midrule
        cofe-Net~\cite{shen2020modeling} & 0.911 & 0.766 & \textbf{0.624} & 0.977 & 0.926 & 0.441 & \textbf{0.468} & 0.807 & 0.202 & 0.0904\\
        PCE-Net~\cite{10.1007/978-3-031-16434-7_49} & 0.879 & 0.696 & 0.540 & 0.976 & 0.949 & 0.500 & 0.328 & 0.874 & 0.102 & 0.134\\
        GFE-Net~\cite{li2023generic} & 0.911 & 0.762 & 0.619 & 0.977 & 0.901 & 0.296 & 0.198 & 0.843 & 0.097 & 0.096\\
        CycleGAN~\cite{cyclegan} & 0.885 & 0.718 & 0.580 & 0.975  & 0.895 & 0.356 & 0.201 & 0.785 & 0.085 & 0.063\\
        SCR-Net~\cite{scrnet} & 0.904 & 0.748 & 0.599 & 0.977 & 0.907 & 0.251 & 0.161 & 0.830 & 0.118 & 0.107\\
        I-SECRET~\cite{i-secret} & 0.878 & 0.695 & 0.531 & 0.977 & 0.909 & 0.275 & 0.214 & 0.783 & 0.044 & 0.056\\
    OTTGAN~\cite{9763342} & 0.896 & 0.740 & 0.592 & 0.977  & 0.936 & 0.453 & 0.278 & 0.871 & 0.169 & 0.136\\
        OTEGAN~\cite{zhu2023otre} & 0.908 & 0.764 & 0.623 & 0.977 & 0.952 & 0.500 & 0.330 & 0.881 & 0.230 & 0.162\\
        \midrule
        Ours & \textbf{0.921} & \textbf{0.772} & 0.591 & \textbf{0.981} & \textbf{0.954} & \textbf{0.565} & 0.343 & \textbf{0.888} & \textbf{0.231} & \textbf{0.164}\\
        \bottomrule
    \end{tabular}}
    \label{tab-seg}
    \vspace{-0.3cm}
\end{table*}

\subsection{Experimental Results}

\noindent \textbf{Enhancement over multiple degradation cases.}
A robust enhancement technique should be able to handle all sorts of degradation and artifacts present in poor-quality images. The Fig. \ref{degrade-perf} illustrates the performance of different degradation cases, that generally occur while capturing the Retinal Images. Our approach preserved the finer and thinner blood vessels very well across the spectrum of noises. In the cases of Blur and/or Illumination degradation where the thinner blood vessels are vaguely visible (Col 1, 3, and 5), our approach has improved the visibility of these vessels, aiding the diagnosis. The Context-Aware optimal transport between the quality domains illustrate complete eradication of the sports artifacts (Cols 2, 4, and 5).

\noindent \textbf{Results in degradation experiments.}
For the degradation experiment, we conducted the qualitative evaluation illustrated in Table~\ref{deg-exp}. We used the pre-trained weights to obtain the enhanced images for cofe-Net, GFE-Net, SCR-Net. For the remaining techniques, we trained the models on default settings shared in the official repository. The proposed method outperformed all baseline methods in terms of PSNR on all three datasets. Specifically, the proposed method surpassed the supervised methods by a significant margin in PSNR on EyeQ, DRIVE, and IDRID datasets, respectively. A similar trend was observed for the GAN-based methods: the proposed method surpassed the recently introduced OT-based OTEGAN by 1.28, 11.54, and 3.45 in PSNR on three datasets, proving our claims of efficient OT-guided learning. 

However, we observed a slight SSIM performance drop in the IDRID and DRIVE datasets (which contain more complex lesion and Vessel structures). We anticipate the problem here to be the limited subjects in these datasets. It is evident that our method stood second for both datasets and maintained a very close margin of 0.007 with CycleGAN for the SSIM on the IDRID dataset. Thus, the proposed method still demonstrated reasonably good generalizability to out-of-distribution datasets (see Table~\ref{deg-exp}). We hypothesized that this was credited to the robust contextual feature encoded in the embedding space that better characterizes image quality. We also observed that although the supervised method showed a satisfactory enhancement performance, it was likely to introduce unrealistic structures (see Fig.~\ref{degrade-exp}). Besides, the other GAN-based methods struggled with removing retinal artifacts. Whereas, the proposed method can remove retinal artifacts.

\noindent \textbf{Results in downstream segmentation tasks.}  
To assess the efficacy of the proposed method in diagnostic tasks, we evaluated its performance on Diabetic Retinopathy lesion and Blood vessel segmentation tasks. As shown in Table~\ref{tab-seg}, our method outperformed others in both tasks, achieving the best ROC and PR scores, proving our claim to better preserve the relevant artifacts during enhancement, essential for proper diagnosis.

\begin{figure*}[!t]
  \centering
  \includegraphics[width=0.93\textwidth]{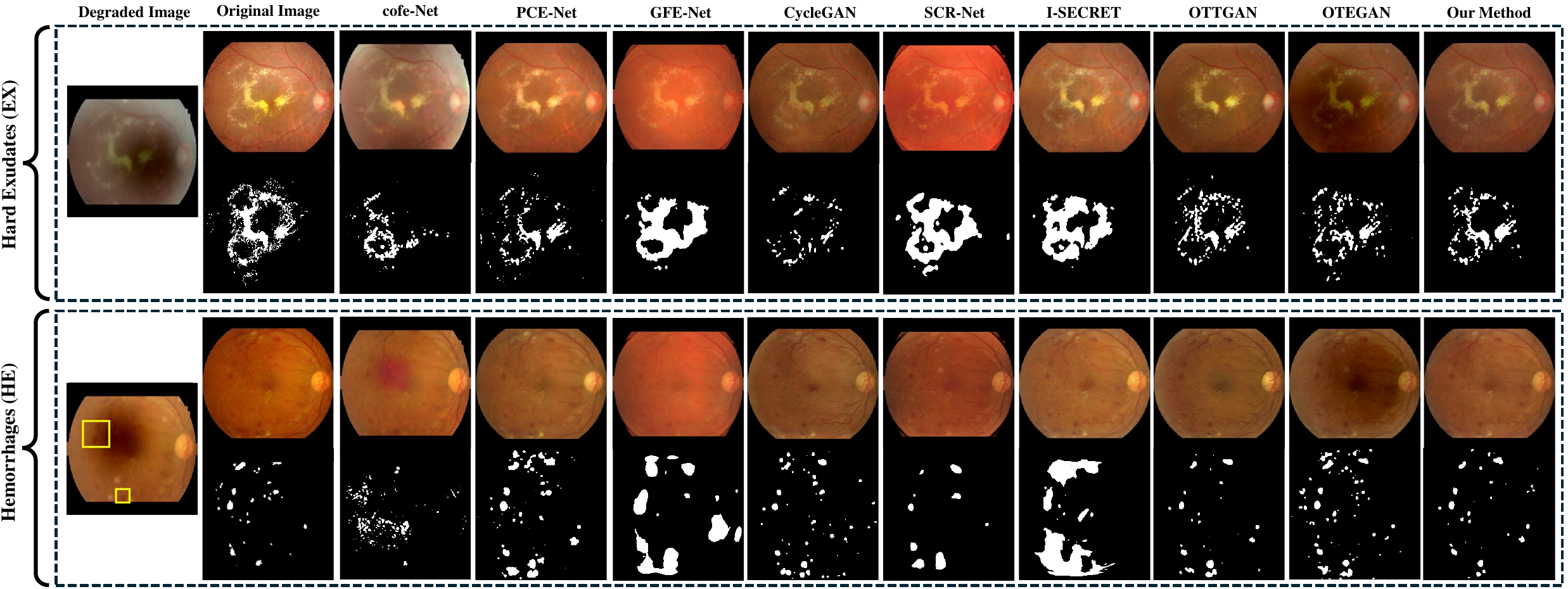}  
  \caption{Lesion segmentation performance over the enhanced images obtained from different methods. We visualized the larger objects, i.e., EX and HE, for better visibility. \textcolor{yellow}{Yellow} boxes highlight the hemorrhage objects that are visually challenging to spot. Compared to other baseline methods, the proposed method leads to the best performance on the downstream lesion segmentation performance, while the baseline methods over-segment or under-segment certain regions.}
  \label{lesion-seg}
\end{figure*}
\begin{figure*}[!t]
  \centering
  \includegraphics[width=0.92\textwidth]{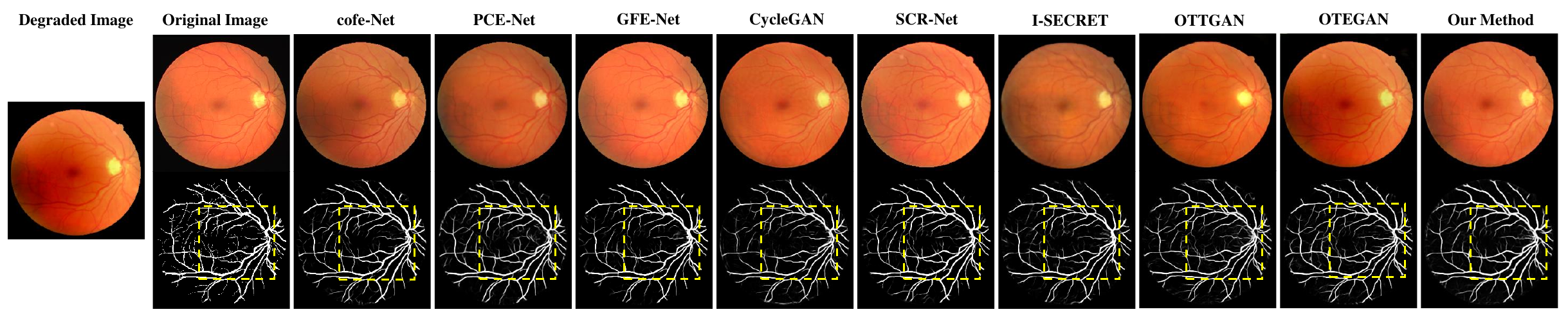}  
  \caption{Vessel segmentation on
the DRIVE dataset. \textcolor{yellow}{Yellow} boxes highlight the preservation of the thinner blood vessels in the denser area of the Fundus. The proposed method leads to the best qualitative results on the downstream vessel segmentation.}
  \label{vessel-seg}
\end{figure*}

We chose two types of lesions, i.e., Hard exudates (EX) and Hemorrhages (HE). We show the qualitative assessment (Fig.~\ref{lesion-seg}) supporting our findings for consistent, superior lesion identification, especially with the HE blocks. 
Despite a performance dip in image enhancement across most methods when untrained on IDRID, our focus remained on lesion preservation. Notably, the other approaches struggled with accurate lesion contouring. Of the two kinds of lesions presented here, EX blocks were easy to locate with the naked eye. But it is another thing to solidly enhance the area. We see that both cofeNet and PCE-Net, even being the supervised method trained over paired data, have yielded incomplete masks. In contrast, the GAN-based techniques classified more areas as lesions. However, the OT-based techniques (OTTGAN, OTEGAN, and Our method) performed the near-accurate prediction. The same is evident from the Quantitative analysis (Table~\ref{tab-seg}).

Hemorrhages are highly indistinguishable when compared to the fundus background. The same is highlighted in Fig.~\ref{lesion-seg} with the yellow box. Considering its delicate nature, most of the methods failed to localize it properly. For instance, I-SECRET often overgenerated the lesions or distorted structures in areas with hemorrhages. However, the context-aware mechanism has overcome this hindrance by outperforming the other methods and achieved top scores for Area under ROC, Precision-Recall, and Jaccard Index. 


We presented the visualization of the Blood Vessel Segmentation task conducted over the DRIVE dataset in Fig.~\ref{vessel-seg}. We outperformed the benchmark methods in the Area under ROC, Precision-recall, and Specificity. The cofe-Net yielded the best Sensitivity value by beating our method by 0.33. Although we see the added advantage of paired data for training in supervised methods, we see cofe-Net missing the thinner blood vessels in the denser region. On the other hand, PCE-Net predicted false vessels (see highlighted region in Fig.~\ref{vessel-seg}). Although the OT-based techniques (OTTGAN and OTEGAN) precisely predicted the finer details in the mask, Our method beat them quantitatively.



\section{Conclusion}

In this study, we propose a Context-Aware Optimal Transport (OT) learning
scheme for enhancing retinal fundus images. For this purpose, we leverage the
earth mover’s distance within the context domain to characterize quality features over extraneous image information. Our findings demonstrated a notable
enhancement over existing benchmarks across three datasets, evidencing the potential of context-aware OT-guided learning in image enhancement. We compared our results with recent state-of-the-art supervised and unsupervised techniques. While our experiments are limited to non-severely damaged datasets, the methodology has the potential for broader application in medical image enhancement, including optical coherence tomography and endoscopy images. We anticipate further exploring the utility of context-aware OT learning in other medical image enhancement applications in future research. 

\section{Acknowledgments} 
\noindent This work was supported by grants from NIH (R01EY032125 and R01DE030286), the State of Arizona via the Arizona Alzheimer Consortium.

{\small
\bibliographystyle{ieee_fullname}
\bibliography{egbib}
}

\end{document}